\documentclass[manuscript,nonacm]{acmart} 
\AtBeginDocument{%
  }

\setcopyright{acmlicensed}
\acmISBN{978-1-4503-XXXX-X/2018/06}

\usepackage{pifont}
\usepackage{float}
\usepackage{tcolorbox}
\usepackage[commandnameprefix=always]{changes}
\begin{document}

\title{{\em M$^3$ QuestionIng}: Multi-modal Multi-span Medical Question Answering}

\author{Anisha Saha}
\affiliation{%
  \institution{Max Planck Institute for Informatics, Saarland Informatics Campus}
  \city{Saarbrucken}
  \country{Germany}}
\email{ansaha@mpi-inf.mpg.de}

\author{Vaibhav Rathore}
\affiliation{%
  \institution{Indian Institute of Technology, Bombay}
  \city{Mumbai}
  \country{India}}
\email{vaibhav.rathore@iitb.ac.in}

\author{Abhisek Tiwari}
\affiliation{%
  \institution{Clinical AI Assistance}
  \city{Gurugram}
  \country{India}
}

\author{Akash Ghosh}
\affiliation{%
 \institution{Indian Institute of Technology, Patna}
 \city{Patna}
 \country{India}}

\author{Sai Ruthvik Edara}
\affiliation{%
  \institution{Yale University}
  \country{United States}}

\author{Sriparna Saha}
\affiliation{%
 \institution{Indian Institute of Technology, Patna}
 \city{Patna}
 \country{India}}


\begin{abstract}
The growing adoption of AI in healthcare, particularly in preventive care, highlights the critical need for accessibility and precision in Medical Question Answering (MedQA). In recent years, significant efforts have been made to develop multi-span medical question-answering systems, where the answer to a query may span multiple sections or paragraphs of a source document. However, existing systems fall short of aligning with real-world scenarios, where source documents often include both textual and visual content, requiring answers to incorporate images for better comprehension. To address this gap, we propose $M^3QAFrame$, a multi-modal, multi-span medical question-answering framework that leverages visual cues to enhance the generation of comprehensive answers drawn from diverse textual and visual spans. The model takes the context, query, and images as input and outputs an answer containing both textual answers and relevant images. The text and image embeddings are processed using a transformer-based architecture to determine the sentence and image relevance. We curate a multi-modal, multi-span medical question-answering ($M^3 QuestionIng$) dataset containing queries, medical contexts, associated medical images, and extractive answers. Additionally, each query-answer pair is labeled with user intent and query type to enhance query and context comprehension. Extensive experiments show that our approach consistently outperforms existing methods across various evaluation metrics.
\end{abstract}




\maketitle

\section{Introduction}
Medical Question Answering (MedQA) plays a crucial role in developing intelligent healthcare assistants by identifying relevant information from diverse contexts. With the lack of a sufficient number of healthcare professionals catering to the needs of a growing population \cite{scheffler2019projecting, lorkowski2021shortage}, the utilization of Artificial Intelligence (AI) based tools has shown promising directions towards reducing medical workloads efficiently \cite{jain2022survey, tiwari2022dr, tiwari2023experience, alsaad2024multimodal, tu2025towards}. Multimodal Medical Question Answering (MMedQA) has gained huge importance over the years due to the availability of medical data in various formats ranging across medical text records, images, video, and audio. The demand for accurate clinical decisions from sensitive medical data aggravates the complexity of this task, and determining a medical ailment from only textual records can be challenging. 
\begin{figure}[t] \centering \includegraphics[width=\columnwidth]{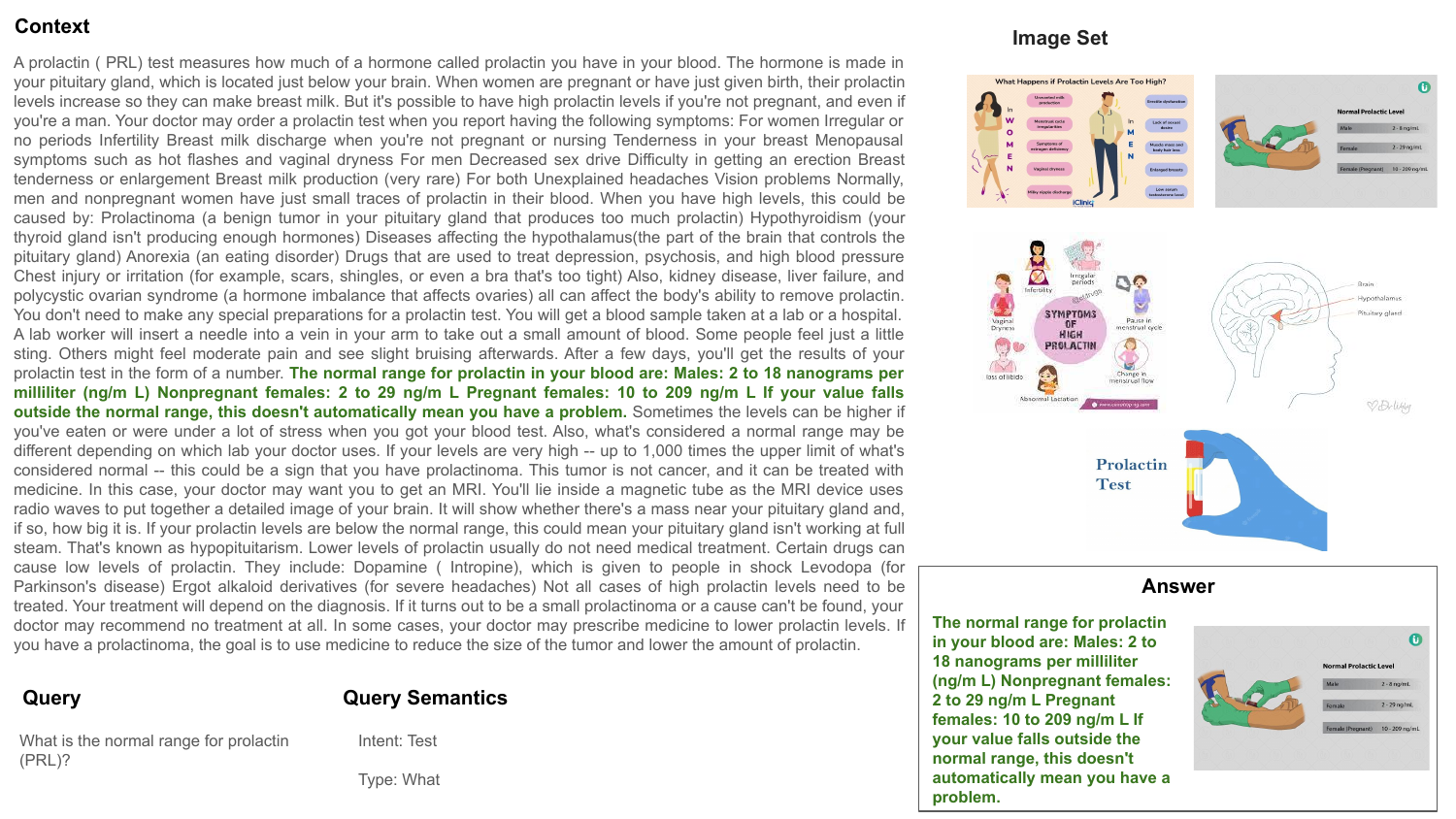} \caption{M³QA processes medical queries by extracting relevant sentences and images from clinical documents and an associated pool of images, thereby incorporating visual information to enhance contextual understanding.} 
\label{iimg} 
\end{figure}
\begin{table}[t]
    \centering
    \caption{Key distinctions between VQA, MedVQA, MedQA, MsQA, and Multi-span Multimodal QA ($M^3$QA). The table highlights different input modalities and answering styles. VQA focuses on general visual QA, MedQA addresses textual QA in medical contexts, MedVQA extends VQA to the medical domain, MsQA supports multi-span answering, and $M^3$QA integrates multimodal fusion with multi-span reasoning.}
    \label{tab:qa_comparison}

    \resizebox{\textwidth}{!}{
    \begin{tabular}{lccccc}
        \hline
        \textbf{Task/Feature} & \textbf{Image Input} & \textbf{Text Input} & \textbf{Medical Domain} & \textbf{Multi-span Answering} & \textbf{Multimodal Fusion} \\
        \hline
        VQA   & \ding{51} & \ding{51} & \ding{55} & \ding{55} & \ding{51} \\
        MedVQA  & \ding{51} & \ding{51} & \ding{51} & \ding{55} & \ding{51} \\
        MedQA   & \ding{55} & \ding{51} & \ding{51} & \ding{55} & \ding{55} \\
        MsQA   & \ding{51} & \ding{51} & \ding{55} & \ding{51} & \ding{55} \\
        \textbf{$M^3$QA (Ours)} & \ding{51} & \ding{51} & \ding{51} & \ding{51} & \ding{51} \\
        \hline
    \end{tabular}}
\end{table}

Multi-span Question Answering (MsQA) refers to answering a query from multiple non-contiguous spans extracted from a given context document. This is unlike single-span QA, where the answer lies in a continuous segment. In practice, 
answering a medical query is a complex process that involves integrating information from multiple sources and modalities. Medical knowledge is communicated through diverse formats including textual descriptions, illustrative diagrams, infographic summaries, and annotated figures, particularly in consumer health education materials such as patient documents, medical encyclopaedias and public health resources. Synthesizing information across these heterogeneous modalities is essential for arriving at a comprehensive answer. In this work, we aim to solve this task of Multi-modal, Multi-span, Medical Question Answering ($M^3QA$). Given a multi-span context $C =(s_1,s_2,...s_n)$ comprising $n$ sentences and a set of $m$ images $I=(i_1, i_2,...,i_m)$ relevant to the context, $M^3QA$ entails identifying the subset of $S$ and $I$ which answer a user query $Q$. In principle, $M^3QA$ is different from Visual Question Answering (VQA) in various aspects as highlighted in Table \ref{tab:qa_comparison}.

In consumer health materials, visual data often plays a critical role in conveying medical information. For instance, illustrative images such as anatomical diagrams, process infographics, and annotated medical figures frequently provide complementary information that is crucial for fully and accurately answering a medical query. Existing approaches heavily rely on textual information or knowledge graphs to guide relevance prediction \citep{ben2019question, shen2020generation}. With the increasing complexity of medical queries, systems need to effectively understand the question’s intent and identify relevant information from diverse modalities, often spanning multiple sentences, contexts, and modalities. QueSemKnow \citep{tiwari2023local} highlighted the importance of leveraging query semantics and external knowledge graphs to enhance multi-span question-answering performance. However, these methods have limitations, particularly in their reliance on static external knowledge, which may fail to address dynamically evolving medical contexts or visual information embedded in medical scenarios. By incorporating image data alongside textual information, we hypothesize that the relevance prediction for medical queries can be significantly improved, leading to more accurate and comprehensive answers. Figure \ref{iimg} shows the relevance of images in answering the given medical query.

To address these issues, we propose a multi-task learning approach $M^3QAFrame$ where a single model is trained to answer questions and identify images relevant to the answer. We hypothesize that the complementary information embedded in images provides a richer representation of the medical context, thereby enhancing the system's ability to identify relevant sentences. Unlike knowledge graphs, which require extensive curation and may suffer from coverage gaps, visual data inherently captures intricate patterns and features specific to medical contexts. To the best of our knowledge, our work is the first to introduce a dataset ($M^3QuestionIng$) and a multi-task framework ($M^3QAFrame$) to solve multi-modal multi-span medical question answering.

\textbf{Research Questions:} In this paper, we investigate the following research questions related to multimodal multi-span medical question answering \textbf{(i)} {\em Does the inclusion of images in the input space help in better identification of the context sentences that form part of the answer?} \textbf{(ii)} {\em Do images contribute unique and complementary information beyond the textual context in medical question answering tasks? }\textbf{(iii)} {\em Do existing vision language models (VLM) show better performance in comparison to models specifically trained for multimodal multi-span medical question-answering?}

\textbf{Contributions:} The key contributions of this work are as follows: 
\begin{itemize}
    \item \textbf{\textit{$\boldsymbol{M^3QuestionIng}$} Corpus} We curate a large-scale semantic information annotated multi-span question answering corpus, $M^3 QuestionIng$, which contains medical contexts, queries, relevant images, intent, and question type for each context-question pair. 
    \item \textbf{\textit{$\boldsymbol{M^3QAFrame}$} Framework} We propose a multimodal multi-task framework $M^3QAFrame$ that integrates image information with text to enhance sentence relevance prediction in extractive medical question-answering, along with identifying images that are relevant to answering those queries.
    \item \textbf{Improved Result}  Through extensive experiments, we demonstrate the effectiveness of our approach, achieving significant improvements in the evaluation metrics over existing methods (approx. 27\%) and over state-of-the-art VLMs (SOTA) (approx. 13\%).
\end{itemize}

\section{Related Works}
MedQA has been a cornerstone of AI-driven healthcare systems, with significant research focused on improving the understanding and retrieval of relevant information from diverse sources. This section reviews related work in three key areas: Medical question answering, Multi-span question-answering and Multi-modal Multi-span question-answering, which are relevant to the present work.

\subsection{Medical Question Answering}
Traditional Medical Question Answering approaches primarily rely on textual data to extract answers. Researchers have employed various techniques, including rule-based question classification \citep{8009118}, knowledge abstraction matching \citep{8982973}, and probabilistic inference on knowledge graphs \citep{goodwin2016medical}. More recently, unified encoder-decoder architectures like UniQA \citep{bae2021question} have been proposed to convert natural language questions into queries. Recent advancements, such as QueSemKnow \citep{tiwari2023local}, introduced a two-phased framework that incorporates query semantics and knowledge graphs to guide multi-span question answering. However, such reliance on static knowledge graphs often limits adaptability to dynamically evolving medical contexts, highlighting the need for alternative sources of auxiliary information. Transformer-based large language models (LLMs), such as BERT \citep{devlin-etal-2019-bert} and RoBERTa \citep{liu2019robertarobustlyoptimizedbert} have been widely employed for natural language understanding tasks and has been adopted to the medical domain in models like BioBERT \citep{lee2020biobert} and ClinicalBERT \citep{huang2019clinicalbert}. Domain-specific LLMs like GatorTron \citep{yang2022gatortron} has been trained to retrieve patient information from unstructured electronic health record, while ClinicalT5 \citep{lu2022clinicalt5} has achieved state-of-the-art performance on natural-language oriented tasks on medical documents including document classification, named-entity recognition and natural language inference. However, they exhibit fundamental architectural and training limitations that render them unsuitable for multi-modal, multi-span question answering. GatorTron, while trained on large volumes of unstructured EHRs, is designed primarily for unimodal text retrieval. It operates over clinical notes and lacks any mechanism to jointly reason over heterogeneous inputs. Besides, the architecture does not support cross-modal attention or grounding and it cannot associate a textual question with visual evidence (e.g., answering a question about tumor from a CT scan image). Similarly, ClinicalT5 achieves strong performance on single-span, text-only tasks that require identifying one contiguous answer within a single modality. In contrast, multi-span question-answering requires the model to simultaneously identify and aggregate multiple non-contiguous text spans and synthesize them into a coherent answer.  These models achieve remarkable accuracy by leveraging pre-trained representations but are limited to processing textual data, thereby excluding critical visual cues present in medical scenarios. 

To enable better clinical reasoning and diagnosis of diseases, multi-modal datasets having radiology images \citep{lau2018dataset}, electronic health records \citep{bae2023ehrxqa} and semantically-labelled images \citep{liu2021slakesemanticallylabeledknowledgeenhanceddataset} has been curated. Multilingual datasets like \citep{matos-etal-2025-worldmedqa} aims to incorporate the knowledge of diverse healthcare environments by introducing 568 QA-image pairs sourced from four countries. Approaches including mixture-of-experts  \citep{jiang2024med} and retrieval-augmented generation \citep{xiammed} has shown to mitigate hallucination in medical QA task. LlaVA-Med \citep{li2024llava} is one of the first multi-modal LLM instruction-tuned on medical data which achieved strong performance on biomedical VQA benchmarks like VQA-RAD \citep{lau2018dataset} and PathVQA \citep{he2020pathvqa}. Following the trend of leveraging pre-trained knowledge of LLMs for medical QA, strong medical VLMs like MedGemma \citep{sellergren2025medgemma}, HealthGPT \citep{linhealthgpt} and LlaVA-Rad \citep{zambrano2025clinically} has been adopted for multimodal clinical reasoning tasks. However these models struggle to handle cross-modal references along long contexts and multiple images in a single conversation. Since retrieval of essential details from a broader context including diverse images is crucial for diagnostic purposes, we cater to this need by developing a multi-modal multi-span medical question answering dataset and propose a method which extracts text segments and images relevant for answering a given medical query.

\subsection{Multi-span Question Answering}
\citet{zhu2020question} introduced the concept of multi-span question answering by contributing a dataset. Multi-span question answering has the potential to significantly improve medical question answering by allowing the extraction of complex and precise information from texts. Traditional QA systems often struggle with questions that require answers from multiple, non-consecutive parts of a document. Most methods emphasize understanding query semantics and retrieving relevant sentences or spans from a given context. To address this, datasets like MASH-QA \citep{zhu2020question} and MultiSpanQA \citep{li2022multispanqa} have been developed to support multi-span question answering. These datasets, along with novel neural architectures like MultiCo \citep{zhu2020question}, can capture the relevance among multiple answer spans and form accurate answers to complex questions. 
\subsection{Multi-modal Multi-span Question Answering}
Recent studies have highlighted the importance of multimodal question answering, where models can reason across multiple modalities, including text, tables, and images. For example, \citet{talmor2021multimodalqacomplexquestionanswering} introduced the MultiModalQA dataset, which requires joint reasoning over text, tables, and images to answer complex questions. \citet{sun2023multimodalquestionansweringunified} proposed a novel multimodal question answering framework that unifies multiple information extraction tasks into a unified span extraction and multi-choice QA pipeline, demonstrating significant improvements over the state-of-the-art baselines. \citet{10535103} proposed a three-staged framework which encompasses unified knowledge representation, context retrieval, and answer generation followed by contextual diversity training to improve model robustness by including distractor documents as negative contexts during training. 

Although significant progress has been made in text-based MedQA, single-span QA and multimodal learning, combining text and image modalities for MsQA remains underexplored. Existing approaches fail to leverage the complementary nature of visual data in answering complex medical queries. This work addresses these gaps by introducing a multimodal multi-span medical question-answering dataset $M^3QuestionIng$ and a multimodal framework $M^3QAFrame$. Our framework accepts large textual contexts and multi-image inputs (unlike most models which lack either of this ability), answers a query that can span across multiple segments of the context, accompanied with visual insights through relevant images, thus improving overall user satisfaction. 

\begin{figure}[t] \centering \includegraphics[width=0.7\columnwidth]{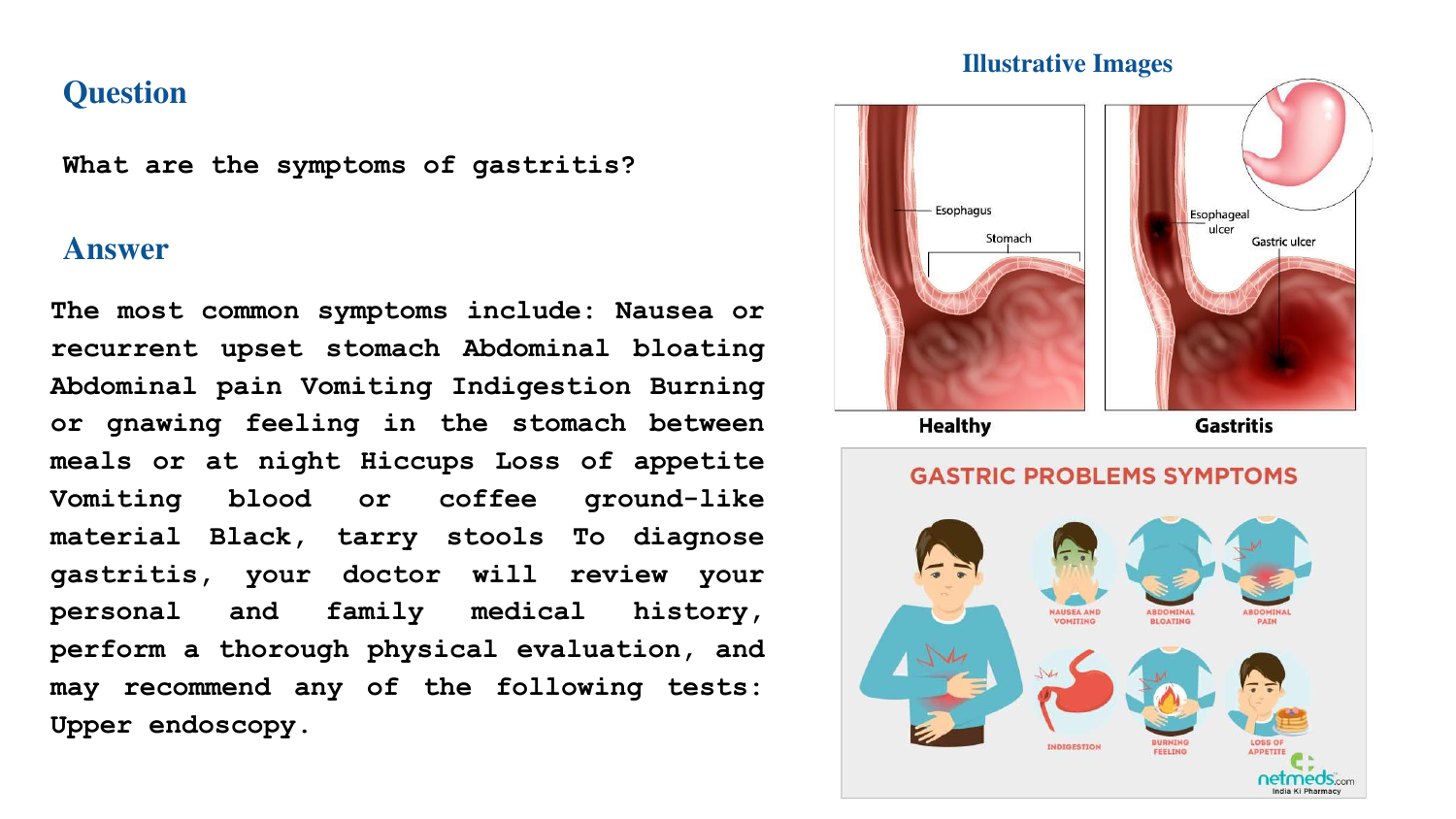} \caption{Infographic-style images enhance interpretability, reflecting real-world patterns in consumer health communication.} 
\label{dataset_image} 
\end{figure}

\section{Dataset} 
We conducted a comprehensive review of existing medical question-answering datasets, and our findings are summarized in Table \ref{DSS}. Datasets such as SLAKE \cite{liu2021slakesemanticallylabeledknowledgeenhanceddataset}, VQA-RAD \cite{lau2018dataset}, and PMC-VQA \cite{zhang2023pmc} are designed for textual, visual, or multimodal question answering, where each query is typically answered with a single, continuous text passage or relevant images. However, real-world medical queries often require information dispersed across multiple text segments and images. Despite our investigation, we did not identify any dataset that supports multi-span, multimodal question answering. To address this gap, we introduce $M^{3}QuestionIng$ — a novel dataset specifically designed for multi-modal, multi-span medical question answering. $M^{3} QuestionIng$ includes structured medical queries, query intents, query types, textual contexts, and associated medical images, providing a more realistic and comprehensive resource for advancing medical question-answering systems. The images in $M^{3} QuestionIng$ illustrate how infographic-style visuals enhance interpretability and engagement, reflecting real-world patterns in consumer health communication where text and visuals jointly support comprehension. The role of these images is not diagnostic but didactic—to help users visually understand key ideas described in the answers. This aligns with the consumer health orientation of the QueSeMSpan dataset \citep{zhu2020question}, where queries are typically posed in layperson language (e.g. “How does sunscreen protect the skin?”). To stay true to this setting, we selected open-source medical illustrations and conceptual diagrams (as shown in Figure \ref{dataset_image}) that explain physiological processes, preventive measures, and general health concepts in an accessible manner.

\textbf{\textit{$\boldsymbol{M^3 QuestionIng}$}}: Developing a medical dataset is a resource-intensive and sensitive process. Therefore, we opted to build upon an existing benchmark dataset by incorporating the missing features necessary to address the identified gap. We identified the MASHQA dataset, which provides multi-span question-answer pairs but is restricted to textual information. To overcome this limitation, we extend QueSeMSpan \citep{tiwari2023local} by introducing both textual and visual question-answering components, thereby creating a more comprehensive dataset for medical question answering. 
\begin{table}
    \centering
    \caption{Comparison of existing multi-span medical question-answering datasets.}
    \label{DSS}
    \resizebox{0.6\textwidth}{!}{
    \begin{tabular}{lcccccc}
        \hline
   \textbf{Dataset} & \textbf{\#QA} & \textbf{Context} & \textbf{Intent} & \textbf{QA Type} & \textbf{Images} \\ \hline
     HealthQA \citep{zhu2019hierarchical}   & 8K & No & No & Ranking & No  \\
     MedQuaD \citep{ben2019question}        & 47K & No & No & Ranking & No \\
     Medication \citep{abacha2019bridging}  & 690 & No & Yes & Ranking & No \\
     MASH-QA \citep{zhu2020question}        & 35K & Yes & Yes & Extractive & No \\
     QueSeMSpan \citep{tiwari2023local}     & 34.8K & Yes & Yes & Extractive & No \\
     M$^{3}$QuestionIng (Ours)              & 3K & Yes & Yes & Extractive & Yes \\  
    \hline
    \end{tabular}}
\end{table}
QueSeMSpan comprises healthcare queries sourced from the WebMD platform \footnote{\url{https://www.webmd.com/}}, encompassing a diverse range of consumer health topics. These queries are answered by medical practitioners with relevant domain expertise, ensuring reliability and accuracy. Each data instance includes a query, query type, intent, and a context-based answer. To enhance the dataset with visual information, we selected 100 data samples randomly from QueSeMSpan and assigned a medical professional to add five relevant images per context. The process involved the doctor first reviewing the context (a set of paragraphs) to identify key concepts that could be better explained with visual aids. Next, the doctor sourced illustrative images corresponding to these key concepts. The tagged images serve as supplementary material to the textual context, enriching the multimodal nature of the dataset. We ensured that all included images are open-source and free of copyright restrictions. 

To scale up the data annotation process, we employed three biology graduates to collect relevant images, tag them with the corresponding contexts, and mark the appropriate images within the answers based on the query requirements. The following guidelines were provided to the annotators to ensure consistency, accuracy, and relevance during the annotation and creation process: \\
\begin{table}[t]
    \centering
    \caption{Statistics of the {\em M$^3$QuestionIng} dataset}
    \label{stats}
    \begin{tabular}{lcc}
    \hline
    \textbf{Entries} & \textbf{Value}\\
    \hline
         \# of samples & 3012\\
         \# of questions annotated with image & 2392 \\
         \# of intents & 11\\
         \# of query types & 12\\
         Avg. context length (in words) & 686\\
         Avg. answer length (in words) & 67 \\
         Avg. image per context & 4.78\\
    \hline
    \end{tabular}
\end{table}
\textbf{Step 1: Context Comprehension} Understanding the context is crucial for accurate annotation. You should thoroughly read the context to identify critical terms, conditions, or concepts that could benefit from visual support

\begin{itemize}
    \item First, carefully read the entire context to grasp the medical information presented.
    \item Identify key terms for each paragraph in the context.
\item Select the key terms that may benefit from visual representation.
\end{itemize}
\textbf{Step 2: Key Concept Identification} Key concept identification helps in selecting images that enhance understanding. This step involves pinpointing important medical terms and concepts within the context.
\begin{itemize}
    \item Identify key terms, conditions, or concepts presented in the document. 
    \item Select a subset of the concepts that require visual aids. 
    \item Highlight the terms that can be better understood with images.
\end{itemize}
\textbf{Step 3: Image Collection and Tagging with Context} This step involves collecting and tagging images based on the identified concepts. 
\begin{itemize}
    \item Images should illustrate the highlighted concepts within the context that enhances understanding.
    \item Each context is tagged with five images.
    \item Only use images from verified, open-source databases to avoid copyright issues.
    \item Ensure image selection is contextually accurate and medically relevant.
\end{itemize}
\textbf{Step 4: Image Tagging with Query-Answer Pair} Accurate tagging ensures images support the query-answer pairs effectively. This step involves linking images directly with relevant query-answer segments.
\begin{itemize}
    \item Tag images to the respective query-answer pairs based on context relevance.
    \item Ensure that images are correctly positioned to aid in answering the queries.
    \item Each answer can be associated with one to a maximum of five images. 
\end{itemize}

\textbf{Step 5: Verification and Correction} In the final step, the medical professional reviews the annotations to ensure accuracy, consistency, and medical relevance. This step is crucial to maintaining the dataset’s quality.

\begin{itemize}
    \item The medical professional verifies that the images accurately represent the medical concepts described in the context.
    \item The individual reviews image-query instances where annotator agreement falls below a defined threshold to ensure accuracy. The professional identifies and corrects any errors or inconsistencies, ensuring the dataset is reliable for medical applications.
    \item We removed the context points with fewer than five images, and query-answer pairs without relevant images were removed to maintain uniformity.
\end{itemize}

A sample dataset was created by medical professionals to serve as training material. We then randomly selected around 3,000 samples from QueSeMSpan and instructed the annotators to collect context-relevant images and assign them to the corresponding query-context pairs. To assess the consistency of annotations, we calculated the inter-annotator agreement using the kappa coefficient, which yielded a value of 0.81 — indicating substantial agreement among the annotators. Upon completion of the annotation process, the dataset was reviewed by a medical professional to validate the annotations and make corrections where necessary. This verification step ensured the reliability of the dataset while maintaining consistency and relevance in its visual components. The dataset statistics are reported in Table \ref{stats}.

\begin{figure*}[t]
    \centering

    \includegraphics[width=\linewidth]{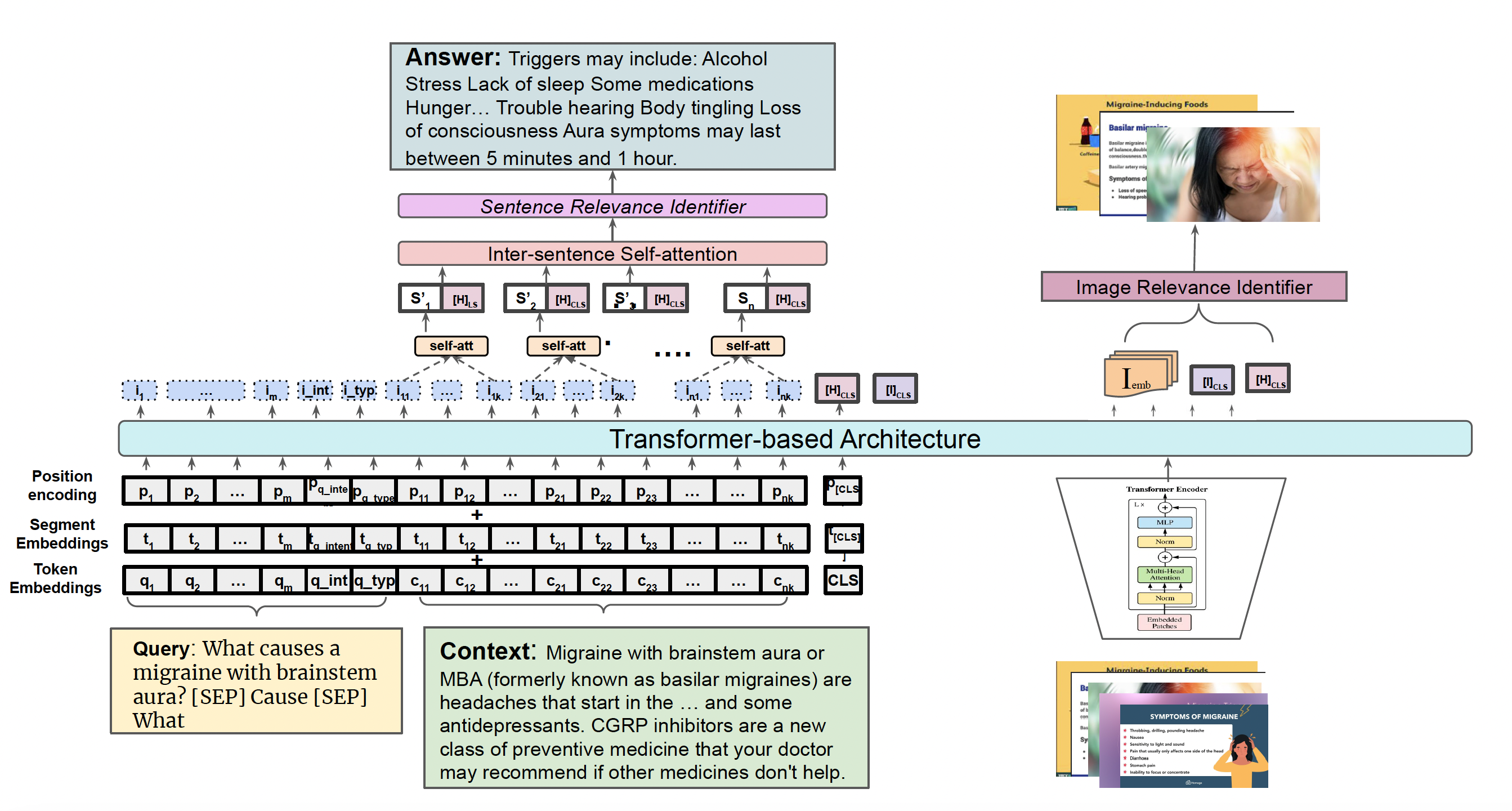}
    \caption{Proposed architecture for multi-modal, multi-span, medical question-answering ({\em $M^3QAFrame$}) model. The system integrates textual context, query, and associated medical images to generate precise and contextually relevant answers. The model leverages visual and textual modalities to address complex queries.}
    \label{model}
\end{figure*}
\section{Methodology}
\textbf{Problem Formulation}: Consider a collection of sentences forming a context $C=<s_1, s_2,...,s_n>$, where $s_j$ represents $j^{th}$ sentence of the context $C$ having $n$ sentences and \textit{m} images $I = <i_{1}, i_{2},..,i_{i_m}>$ associated with the context $C$.  Given a query $Q$, whose answer belongs to the context $C$, multi-modal, multi-span question answering refers to identifying the subset of relevant sentences $A_C=<s_i, s_j,...,s_k>$, contiguous or non-contiguous, and the subset of relevant images $A_I=<i_{a}, i_{b},..,i_{c}>$ which answer the given query. For each query in the dataset, corresponding to a context, there is a unique set of sentences (collection of all the relevant sentences) which answers the query. This excludes the possibility of multiple correct answers for a query.\\

The proposed model architecture (Figure \ref{model}) is inspired by the MSQA framework \citep{tiwari2023local}. We add on a multi-tasking setting where we predict the relevant sentences and images which answer a user query. The model functions in three steps: (i) Text Processing (ii) Image Processing (iii) Fusion and Classification. While our architecture builds on the foundational transformer design (chosen specifically to handle long contexts), it introduces novel enhancements specific to the medical domain: (a) Integration of medical-domain-specific encoders. (b) Customized attention mechanisms and multi-task learning for sentence and image relevance prediction. (c) Focus on complete multi-modal integration both in the input and output. 

\subsection{Text Processing}
The text processing module encodes the query and the context. Given a query $Q = <Question, Intent, Type>$ and a context $C=<s_1, s_2,...,s_n>$, where $s_j$ represents $j^{th}$ sentence of the context $C$ having $n$ sentences,  we concatenate $Q$ and $C$ with a special token ([SEP]) to obtain a single text input. Next, the concatenated text is passed through BiomedBERT \citep{Gu_2021} to obtain the text encodings. Since the text contains dense medical terminologies and references, BiomedBERT is chosen to leverage its biomedical domain-specific pretraining.
\begin{equation}
    T_i = BiomedBERT(Q_i~[SEP]~C_i)
\end{equation}
where $T_i$ is the $i^{th}$ text encoding of the $N$ data instances. Since the downstream objective is to predict relevant context sentences, each $s_{ij}$ ($j_{th}$ sentence of $i_{th}$ context $C_i$) is separated using a special token ([CSEP])
\subsection{Image Processing}
The image encodings are obtained using Vision Transformer \citep{dosovitskiy2021imageworth16x16words} model trained using the DINO method \citep{9709990}. Unlike most models, $M^3QAFrame$ accepts multiple images as input. For justifying the relevance of images in multi-span question answering (RQ1), for all the images $I_j = <i_{i1}, i_{i2},..,i_{iM}>$ associated with a given context $C_i$, we obtained the average of all the image embeddings:
\begin{equation}
    i_{im}^{'} = ViT(i_{im})
\end{equation}
\begin{equation}
    I_{i}^{rel} = mean(i_{i1}^{'}, i_{i2}^{'},...,i_{iM}^{'})
\end{equation}
However, for the image classification task, we obtain embeddings for individual images, concatenate them using a special token ([ISEP]), followed by a classification token ([ICLS]).
\begin{equation}
    I_{i}^{cls} = i_{i1}^{'}~[ISEP]~i_{i2}^{'} ~ [ISEP]...i_{iM}^{'}[ICLS]
\end{equation}
where $M$ denotes the number of images present for the context $C_i$ and $M$ varies across samples.
\subsection{Fusion and Classification}
To obtain the answer to a query, we need to focus on the following key components: (a) Query understanding (b) Textual context understanding (c) Image context understanding  (d) Global context provided by text+image. Since the answer should also consist of the relevant images, the problem we solve here is two-headed. The following stages help us capture intricate details from both modalities.

\textbf{Self Attention Layer}: Since the contexts are of varied lengths, we apply a self-attention \citep{10.5555/3295222.3295349} to obtain vectors of a fixed length and understand inter-query-context and inter-context semantics, which is calculated as follows:
\begin{equation}
    h_{ij} = w_{s}tanh(W_{s}.C_{ij}^{'})
\end{equation}
\begin{equation}
    att_{ij} = Softmax_{i}(h_{ij})
\end{equation}
\begin{equation}
    S_{i}' = \sum_{j=1}^{j=k} att_{ij}.C{j}^{'}
\end{equation}
Here, $w_{s}$ and $W_{s}$ are learnable parameters, $C_{ij}^{'}$ denotes the encoded representation of $jth$ word of $ith$ sentence of the context and $S_{i}'$ indicates the attended hidden representation for the $ith$ sentence for the context $C_{i}$.

\textbf{Inter-sentence self-attention}: Since the number of sentences that are part of the final answer is very small in number with respect to the total number of sentences in the context, the traditional method of attention-weight calculation using softmax is highly likely to suffer from skewness. To mitigate this, we calculate sparsified inter-sentence self-attention $\alpha$-entmax \citep{peters-etal-2019-sparse} as follows:
\begin{equation}
    si\_sa_{ij}=w_{s}.tanh(W_{s}.S_{ij}')
\end{equation}
\begin{equation}
    \beta_{ij} = f_{s}(si\_sa_{ij})
\end{equation}
\begin{equation}
    S_{i}'' = \sum_{j=1}^{j=k} \beta_{ij}S_{ij}'
\end{equation}
\begin{equation}
    f_s = ReLU[(\alpha - 1).a - \tau]^{1/\alpha - 1}
\end{equation}
Here $f_s$ is a sparse attention function, which unlike traditional softmax pays attention towards relevant sentences over irrelevant ones. $\beta_{ij}$ denotes the attention weight of the $ith$ sentence with respect to the subject $jth$ sentence and $S^{'}_{ij}$ is the $jth$ word of $S_{i}^{'}$. The [ICLS] token which captures the image information is concatenated with each of the attended sentences to utilize the image information in sentence relevance prediction.

\textbf{Sentence and Image Relevance Identification:}
The attended sentence vectors are now passed through a feed forward neural network. The final layer consists of a softmax function which assigns labels to whether a sentence is relevant or not. 
\begin{equation}
    y_{i} = softmax(W_{o}.S_{i}'' + b_{o})
\end{equation}  

Similarly, the image vectors concatenated with the [CLS] token encoding the text information are passed through a feed forward network followed by a softmax layer. Here, $W_{0}$ and $b_{o}$ are the weight and bias term, respectively. The model employs binary cross-entropy loss to backpropagate the discrepancy between the actual data and the model’s predictions. It is calculated as follows:
\begin{equation}
    \scalebox{0.9}{$loss = - \sum_{j=1}^{j=N} \sum_{i=1}^{i=n}[y_{i}^{(j)}log( \hat{y_{i}}^{(j)}) + (1-y_{i}^{(j)})log(1-\hat{y_{i}}^{(j)})]$}
\end{equation}

A similar loss is calculated for image classification. Here, $y_i^{(j)}$ and $\hat{y_i}^{(j)}$ are the predicted and true probabilities, respectively, for the $i^{th}$ sentence (image) of the $j^{th}$ sample being considered as answer (image relevant to the answer), $N$ is the total number of samples and $n$ is the number of sentences (images) in the respective samples. 

To ensure effective learning, the total loss is formulated as a weighted combination of the sentence relevance loss (\( \mathcal{L}_{sent} \)) and the image relevance loss (\( \mathcal{L}_{img} \)), where \( \lambda_s \) and \( \lambda_i \) are the respective weighting coefficients, satisfying \( \lambda_s + \lambda_i = 1 \). The total loss function is given by:

\begin{equation}
\mathcal{L}_{total} = \lambda_s \cdot \mathcal{L}_{sent} + \lambda_i \cdot \mathcal{L}_{img}
\end{equation}

\section{Experimental Setup}
The $M^{3}QAFrame$ model was trained for 25 epochs on RTX 2080 Ti GPU, with each epoch taking approximately 4 hours. The train, validation, and test set had a division of 80\%, 10\%, and 10\%, respectively. 
We used grid search to obtain the optimal set of hyperparameters: batch size (8), learning rate $\alpha$ (0.00003), and optimizer (Adam). We observed that choosing \( \lambda_s = 0.7 \) and \( \lambda_i = 0.3 \) helped the model learn better justifying the fact that extracting relevant sentences from a large context pool is inherently more challenging than selecting a smaller subset of relevant images, justifying the chosen weighting scheme. The reported results are averaged for runs over five random seeds. We provide ablation studies 

\hspace{-0.35cm}\textbf{Baselines}: We perform a comparative analysis of our model's performance over the existing models which have been known to perform well for the task of MsQA in the past. These include BERT \citep{devlin-etal-2019-bert}, RoBERTa \citep{liu2019robertarobustlyoptimizedbert}, TANDA \citep{Garg2019TANDATA}, MultiCo \citep{zhu2020question} and QueSemKnow \citep{tiwari2023local}. Note that since this is the first work on MsQA with both multi-modal inputs and outputs, the above baselines we compare against are unimodal. Due to unavailability of a multi-modal baseline, we evaluated the performance of XLNet \citep{10.5555/3454287.3454804} and Longformer \citep{beltagy2020longformer} by switching places with BiomedBERT and keeping the rest of the model pipeline constant.
\begin{table}[t]
    \centering
    \caption{Performance comparison of various MsQA models with respect to $M^{3}QAFrame$}
    \label{textimage}
    \scalebox{0.7}{
    \begin{tabular}{lccc}
        \hline
   \textbf{Model} & \textbf{Accuracy} & \textbf{F1-Score} & \textbf{Exact Match}\\
        \hline
     BERT  & 24.82 & 25.21 & 8.89 \\
     RoBERTA  & 26.89 & 28.65 & 9.40 \\
     TANDA  & 23.48 & 24.44 & 8.95 \\
     MultiCo & 46.21 & 50.81 & 17.80 \\
     QueSemKnow & 58.31 &  55.81 & 21.33 \\  
     XLNet w/o image  &  38.01 & 29.19 & 9.09 \\ 
     XLNet w/ image  &  39.28 & 30.67  & 10.11\\ 
     Longformer w/o image &  61.85 & 67.13 & 61.85\\ 
     Longformer w/ image  & 65.23  & 69.17 & 65.23 \\ 
     $M^{3}QAFrame$ w/o image & 90.51 & 86.10 & 90.51 \\
     \textbf{$\mathbf{M^{3}QAFrame}$ w/ image (Ours)} & \textbf{94.34 (29.11 $\uparrow$)} & \textbf{91.32 (22.15 $\uparrow$)} & \textbf{94.34 (29.11$\uparrow$)}\\
    \hline
    \end{tabular}}
\end{table}
\begin{figure*}[t] 
    \begin{minipage}{0.47\textwidth} 
        \centering
        \includegraphics[width=\linewidth]{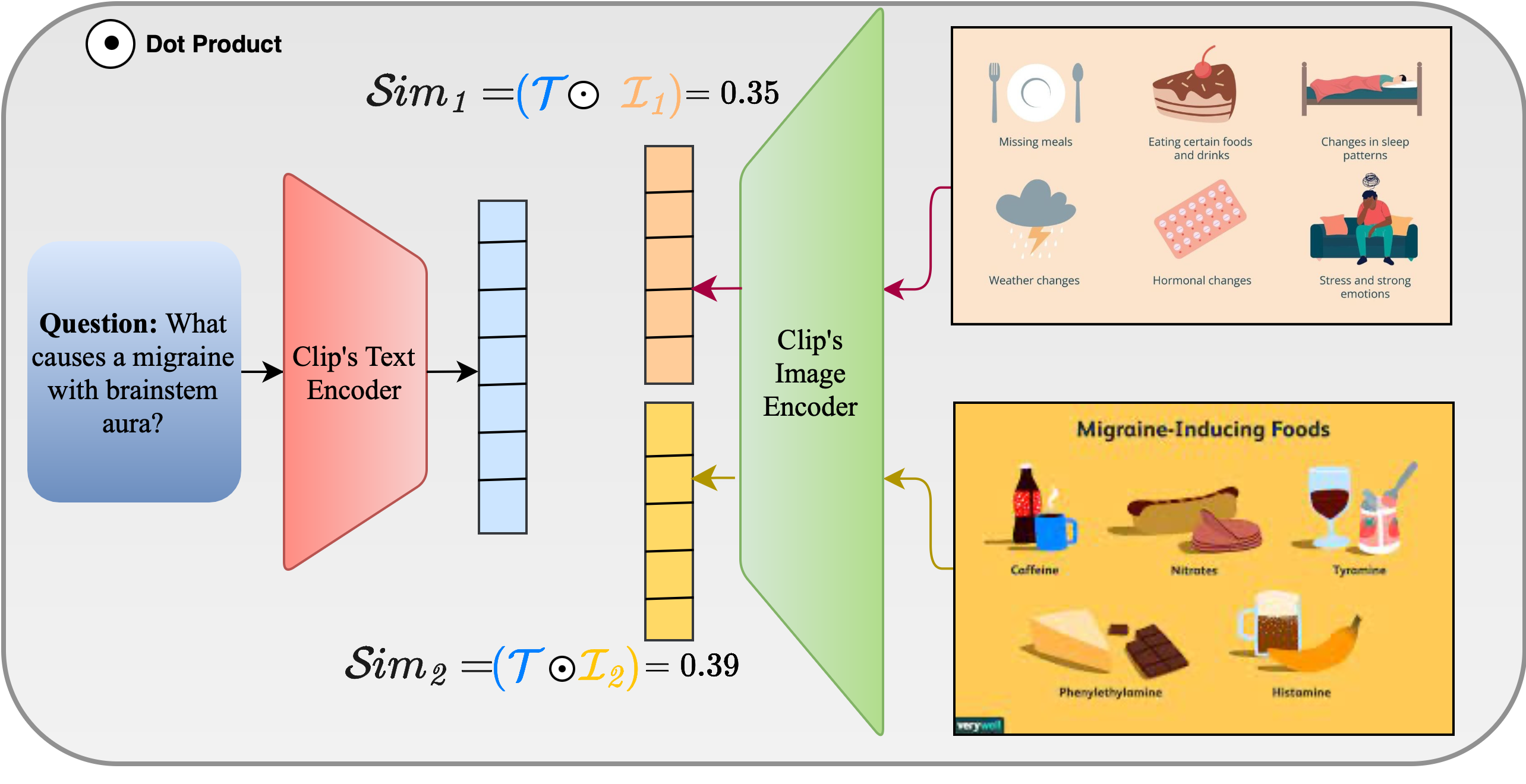}
        \caption{Evaluating the role of images in medical QA using CLIP similarity. Here, visual and textual contexts are aligned to measure semantic relevance and complementary information, informing the development of more accurate multimodal MedQA systems.}
        \label{example1}
    \end{minipage}
    \hfill 
    \begin{minipage}{0.48\textwidth} 
        \centering
        \includegraphics[width=\linewidth]{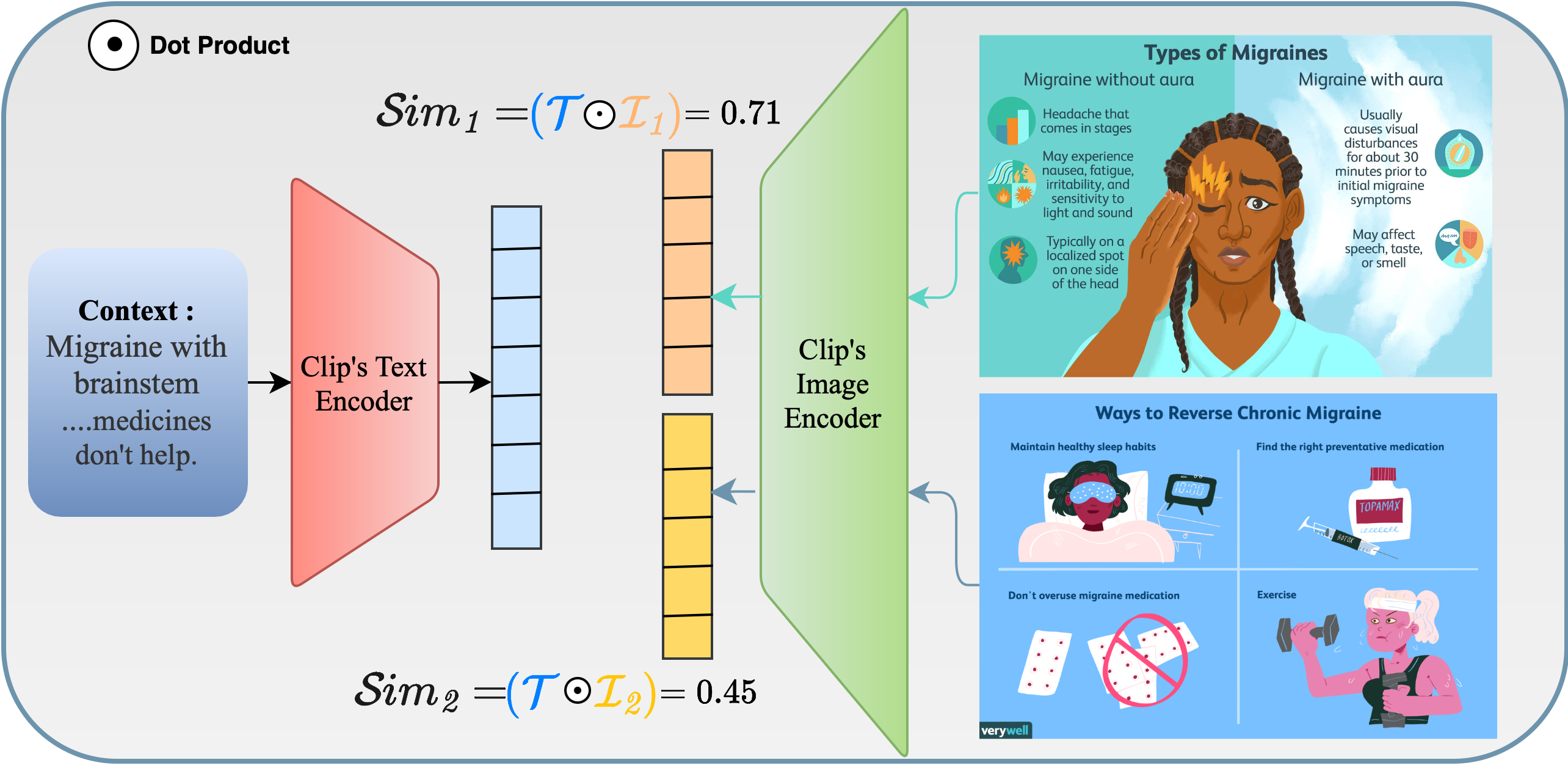}
        \caption{Analyzing textual-image relationships in medical QA using CLIP encoders. Semantic similarity scores (\(Sim_1\) and \(Sim_2\)) reveal how images complement or reinforce textual details, guiding improved multimodal MedQA approaches.}
        \label{example2}
    \end{minipage}
    \captionof{figure}{Comparison of CLIP similarity evaluation (left) and textual-image relationship analysis (right) in medical QA. These figures collectively highlight the role of visual and textual contexts in developing robust multimodal MedQA systems.} 
    \label{combined_examples}
\end{figure*}

\section{Results and Discussions}
Since sentence and image-relevance identification is a multi-label classification task, we use Accuracy, F1-score and Exact Match to evaluate the performance of the models for multimodal multi-span question answering. For Accuracy and F1-Score, we use the standard Scikit-Learn implementation of \textit{accuracy\_score} and \textit{f1\_score $($average=`macro'$)$} respectively. For Exact Match, we use the \textit{exact\_match} implementation of \textit{evaluate} library. The score is 1 if two sentences are exactly the same and 0 otherwise. For the reference and predicted answers, the score is averaged across exact matches for individual sentences.  We report the performance only on $M^3QuestionIng$ dataset as it is the first of its kind with no prior dataset having the same structure and components.A mathematical formulation of calculating the above scores are as below:

Let a query $Q_i$ have a set of \textit{reference answer sentences},
\begin{equation}
\mathcal{A}_i = \{a_i^1, a_i^2, \ldots, a_i^{k_i}\}
\end{equation}

and a set of \textit{predicted answer sentences},

\begin{equation}
\hat{\mathcal{A}}_i = \{\hat{a}_i^1, \hat{a}_i^2, \ldots, \hat{a}_i^{j_i}\}
\end{equation}

where $k$ is the number of answer sentences (spans) for question $q_i$. For each individual sentence pair, the binary accuracy (and exact match) is defined as:

\begin{equation}
score_i = 
\begin{cases} 
1 & \text{if } a_i^l = \hat{a}_i^m \\ 
0 & \text{otherwise} 
\end{cases}
\end{equation}
The per-query score is the \textit{average exact match across its $k_i$ answer sentences}:

\begin{equation}
\text{EM}_i = \text{Accuracy}_i = \frac{1}{k_i} \sum_{p=1}^{k_i} score_p
\end{equation}

The final dataset-level Exact Match / Accuracy score over all $N$ queries is:

\begin{equation}
\text{EM} = \text{Accuracy} = \frac{1}{N} \sum_{i=1}^{N} \frac{1}{k_i} 
\sum_{p=1}^{k_i} score_p
\end{equation}

Precision counts the fraction of predicted sentences that exactly match a reference sentence, and Recall counts the fraction of reference sentences that were exactly predicted. These are computed over matched sentences as:

\begin{equation}
\text{Precision}_i = \frac{\sum_{p=1}^{k_i} score_p}{|\hat{\mathcal{A}}_i|}
\qquad
\text{Recall}_i = \frac{\sum_{p=1}^{k_i} score_p}{|\mathcal{A}_i|}
\end{equation}

where $|\mathcal{A}_i|$ and $|\hat{\mathcal{A}}_i|$ are the number of reference 
and predicted sentences respectively for query $Q_i$.

The per-question F1 score is:

\begin{equation}
\text{F1}_i = \frac{2 \cdot \text{Precision}_i \cdot \text{Recall}_i}
{\text{Precision}_i + \text{Recall}_i}
\end{equation}

The dataset-level F1 score over all $N$ queries is:
\begin{equation}
\text{F1} = \frac{1}{N} \sum_{i=1}^{N} \text{F1}_i 
\end{equation}

\hspace{-0.35cm}\textbf{RQ1: Does inclusion of images in the input space help in better identification of the context sentences which form the answer to a query?} The results for sentence-relevance prediction in both the unimodal and multi-modal setting are shown in Table \ref{textimage}. We observe that the inclusion of images demonstrates a significant improvement not only in our proposed model's performance but also in the case of XLNet and Longformer. It establishes strongly the fact that additional images provide additional information which are utilized by the model to obtain better predictions. 


\hspace{-0.35cm}\textbf{RQ2: Do images contribute unique and complementary information beyond textual context in medical question answering tasks?} . To investigate this, we tried to quantify this additional information using CLIP similarity score \citep{hessel-etal-2021-clipscore}. By analyzing these similarity scores, we can assess how much value images add to answer a query. Two types of similarity are calculated:
\begin{itemize}
    \item {\em Context Images \& Question}: To measure whether the image is semantically relevant to the question. In Figure~\ref{example1}, two context images achieve cosine similarities of 0.35 and 0.39 with the question, providing moderate yet useful visual cues that enhance the model’s predictions.
    \item {\em Context Images \& Context Text}: To measure whether the image adds information complementary to the textual context.   Figure~\ref{example2} presents a scenario where one image closely aligns with the context text, showing a high similarity of approximately 0.71 and thus contributing less novel information. Another image in the same context exhibits a lower similarity of about 0.45, indicating that it provides unique complementary details beyond the textual content.
\end{itemize}

 This similar trend across images in the dataset confirms that these images frequently offer complementary information, particularly when their semantic alignment with the question or context text differs. In cases demanding domain-specific visual reasoning, these additional visual cues improve model performance and lead to more accurate and reliable medical question answering. Thus we conclude that images associated with a multi-span document provide additional context that cannot be derived from text alone.

\hspace{-0.35cm}\textbf{RQ3: Do existing vision language models (VLM) show better performance in comparison to models specifically trained for multi-modal multi-span medical question-answering? } We performed zero-shot and few-shot evaluations on both domain-specific and general purpose state-of-the-art (SOTA) VLMs like Uni-MedCLIP \citep{khattak2024unimed}, MMEmbed \citep{lin2024mm}, VLM2Vec \citep{jiang2024vlm2vec}, LLaVA \citep{liu2024visual} and GPT-4o \citep{openai2024gpt4ocard}, and report average scores of both the settings due to very low deviation between them. In addition we also fine-tune Qwen2.5 VL \citep{bai2025qwen25vltechnicalreport} and MedGemma \citep{sellergren2025medgemma} for a fair comparison against our proposed architecture trained on $M^3QuestionIng$. The prompts used are provided in the Appendix. \newline
\begin{figure*}[t]  
    \centering
    \includegraphics[scale=0.42]{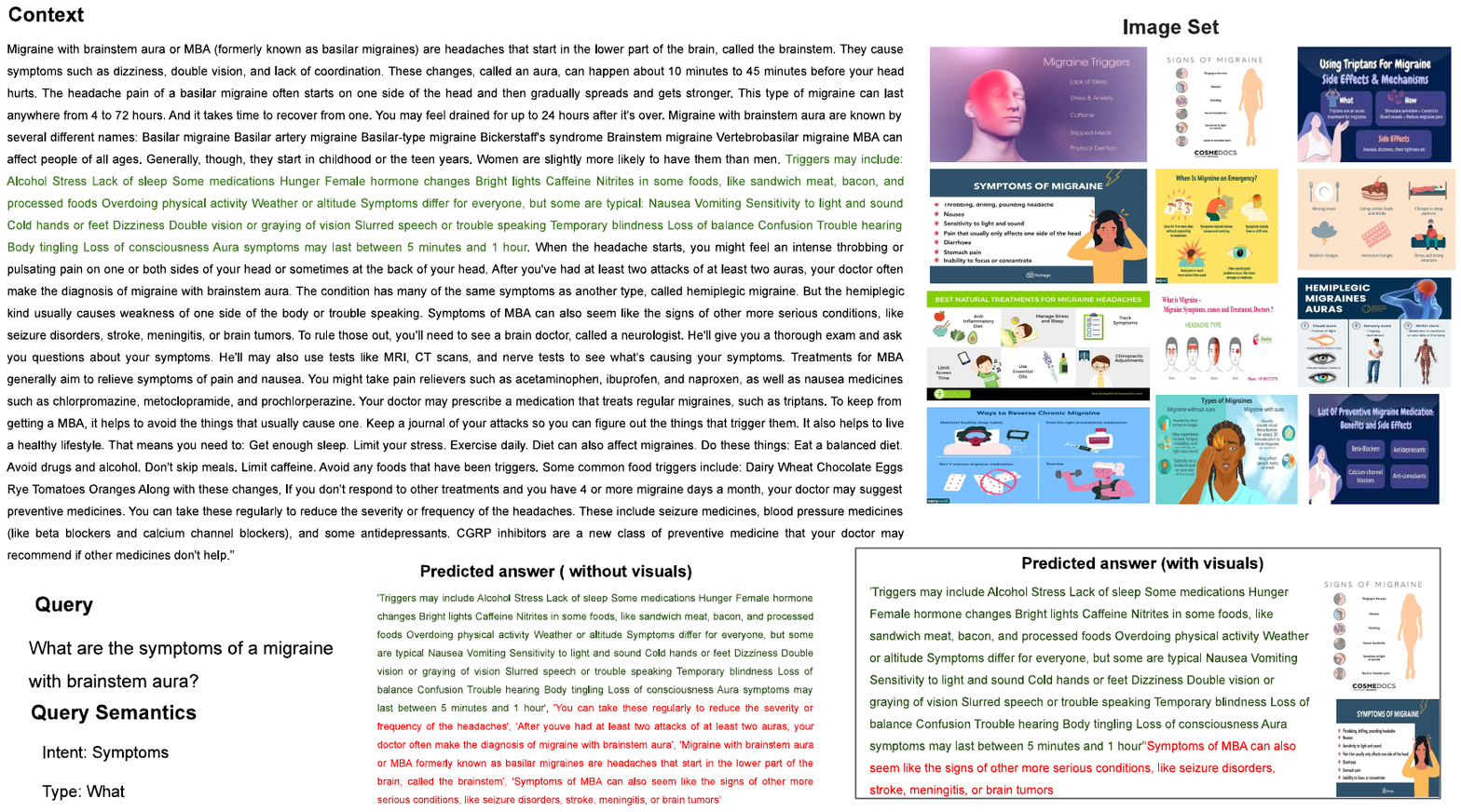}  
    \vspace{-3em}
    \caption{Illustration of the importance of visual in Multi-span Question Answering}
    \label{visual}
\end{figure*}
\begin{table}[t]
    \centering
    \caption{Comparison of performance between our proposed model $M^3QAFrame$ and SOTA VLMs in the MsQA task.}
    \label{imgrel}
    \resizebox{0.7\linewidth}{!}{
    \begin{tabular}{lccc}
        \hline
   \textbf{Model} & \textbf{Accuracy} & \textbf{F1-Score} & \textbf{Exact Match}\\
        \hline
    Uni-MedCLIP  & 12.74 & 22.60 & 11.78 \\
    MMEmbed  & 52.15 & 66.67 & 34.59 \\
    VLM2Vec  & 28.05 & 43.51 & 27.84 \\
    LLaVA  & 40.56 & 56.32 & 40.87 \\
    Qwen2.5-VL \textit{$($finetuned$)$} & 65.38 & 5.07 & 65.38 \\
    MedGemma \textit{$($finetuned$)$} & 39.16 & 55.05 & 39.16 \\
    GPT-4o & 81.2 & 78.5 & 79.5 \\
    $\mathbf{M^{3}QAFrame}$ \textbf{(Ours)} &  \textbf{94.34 (13.14 $\uparrow$)} & \textbf{91.32 (12.82 $\uparrow$)}& \textbf{94.34 (14.84 $\uparrow$)}\\
    \hline
    \end{tabular}}
\end{table}

We exclude comparisons with models like MedVQA models, such as PMC-CLIP \citep{lin2023pmc}, BioViL \citep{bannur2023learning} and LLaVA-Med \cite{li2024llava} because of the following reasons: (i) The length of the contexts in our dataset is huge (Max. words in a context = 2800 ), which cannot be handled by these models. Since, the purpose of our work is answering a query with sentences spanning the entire context, truncating a context to 10-20\% of its actual length defeats the whole purpose. (ii) These models do not accept multiple images as input, unlike our model. (iii) They are trained for different downstream tasks.
Table \ref{imgrel} illustrates the performance gain in $M^3QAFrame$'s thereby establishing the need for fine-tuned deep networks for MsQA over SOTA VLMs. 


\begin{figure}[t]  
    \centering
    \includegraphics[width=0.9\columnwidth]{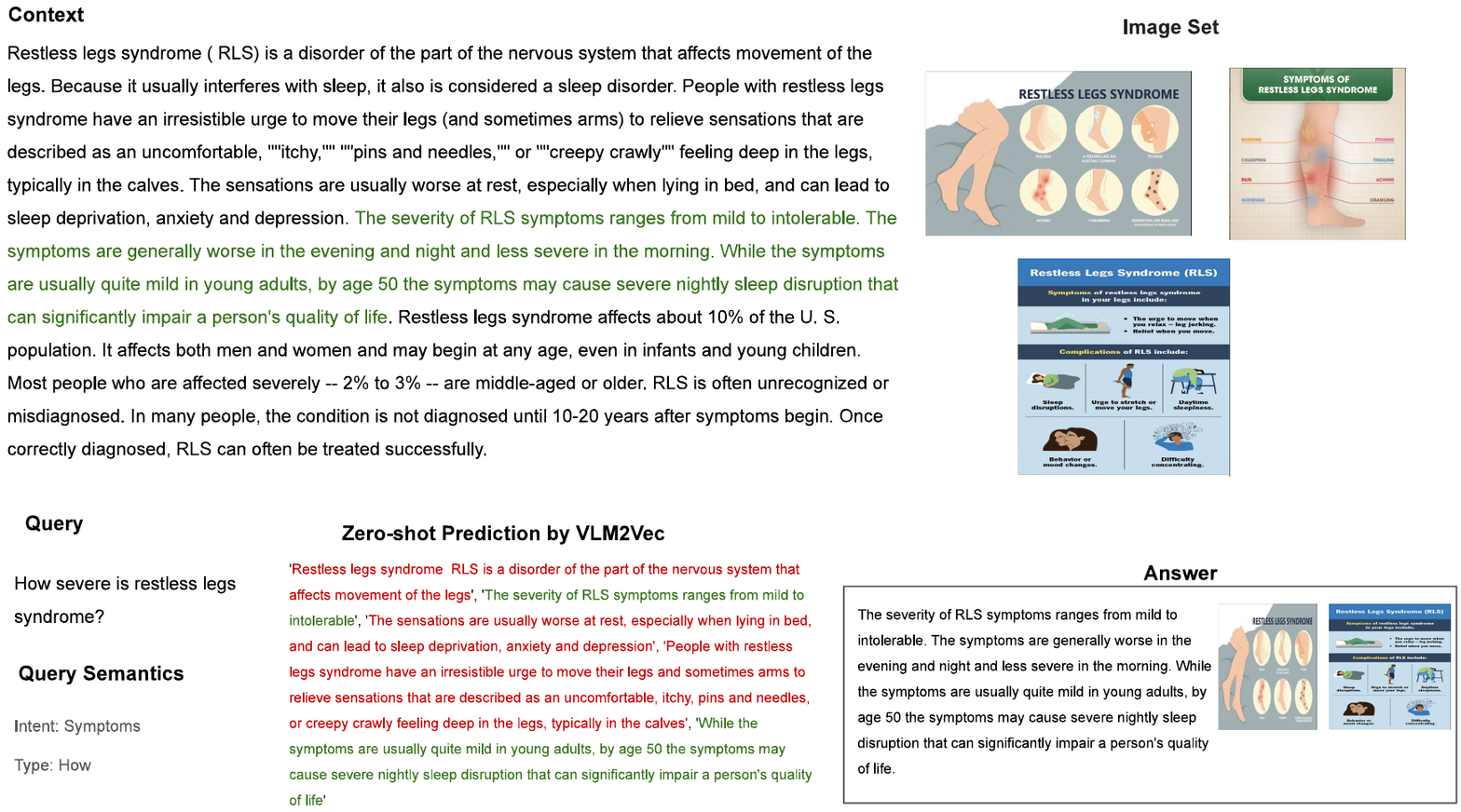}  
    \vspace{-2em}
    \caption{Example of VLM2Vec's performance}
    \label{vlm}
\end{figure}

\begin{table}[t]
    \centering
    \caption{Impact of various elements of  $\mathbf{M^{3}QAFrame}$}
    \label{ablation}
    \resizebox{0.7\linewidth}{!}{
    \begin{tabular}{lccc}
        \hline
   \textbf{Model} & \textbf{Accuracy} & \textbf{F1-Score} & \textbf{Exact Match}\\
        \hline
    $M^{3}QAFrame$ w/o \textit{Intent}  & 91.75 & 87.66 & 91.75 \\
    $M^{3}QAFrame$ w/o \textit{Query} & 92.57 & 88.20 & 92.57 \\
    $M^{3}QAFrame$ w/o \textit{Self-attention}  & 84.14 & 80.37 & 84.14 \\
    \hline
    \end{tabular}}
\end{table}
\section{Case Study and Analysis}\label{casestudy}
To analyze the model's performance in different scenarios, this section details a closer look at the generated outputs, adversarial scenarios where the model fails, and the importance of various building blocks of the model.

\begin{itemize}
    \item \textbf{Impact of images:} One of the major motivations of our work is to study how essential are images for MsQA. We observe a significant improvement in our model's performance when images accompany text as input. An example of this can be found in Figure \ref{visual}.
    \item \textbf{Performance of VLMs:} Figure \ref{vlm} illustrates the subpar performance of VLMs on the $M^3QA$ task. This can be attributed to the lack of similar structured data in the pre-training datasets. For example, the notably low F1 score (5.07\%) observed for the fine-tuned Qwen2.5-VL model can be attributed to the structural mismatch between how the model generates answers and how multi-span answers are evaluated. Qwen2.5-VL, being a generative VLM, tends to produce free-form responses rather than concise discrete answer spans which is a common behavioral tendency of instruction-tuned autoregressive models. As a result, the predicted answer set either contains many sentences that do not exactly match any reference span or are structurally misaligned with the expected multi-span format. This behavior highlights the limitation of adapting VLMs to structured multi-span QA tasks. Besides, these VLMs have not been specifically trained on the downstream task of sentence relevance prediction which is core to $M^3QA$. Additionally, a majority of the open-source VLMs cannot generate images as output.
    
    \item \textbf{Impact of architecture components:} The ablation study in Table \ref{ablation} demonstrates that query type is more essential than query intent for answer extraction. It also highlights the importance of self-attention for the model to develop semantic understanding between the query and the large context, owing to inter-sentence reasoning over a large number of pairs.
\end{itemize}

\section{Limitations and Future Work}
The proposed dataset and data model provide a foundation towards multi-modal, multi-span medical question answering, which has not been explored much so far. Although the dataset has been created using stratified sampling which reduces systematic bias, it might still favor medical topics that are visually representable (e.g., anatomical or dermatological content). Besides, the model architecture could be improved to precisely cater to the needs of domain-specific understanding. In the current work, we do not use a medical domain-specific vision encoder like MedCLIP \citep{Wang2022MedCLIPCL}, which could be an addition to the present model in the future. We also observed that although images provide extra information, the present vision encoder cannot properly identify the embedded texts within the image. Further, we use two different frozen text and vision encoders, respectively, and only learn the succeeding layers. In future, we aim to explore sophisticated modality fusion and alignment techniques like Q-Former \citep{li2023blip2bootstrappinglanguageimagepretraining}. We also look forward to exploit the pretrained knowledge of existing vision-language foundation models and customize them to the task of multimodal multi-span question-answering. Our work currently does not involve an image generation module which could help generate images relevant to the answers for illustrative purposes. Future extensions leveraging transformer-based diffusion models could be integrated into the current architecture to enable generating reference images aligned with multi-span answer contexts.

\section{Conclusion}
In this paper, we try to address the problem of multi-modal multi-span medical question-answering. For a given user query, our approach retrieves a textual answer spanning across multiple segments of a document, along with relevant images that provide a visual insight into the answer, improving user experience. We curated a first-of-its-kind dataset $M^3 QuestionIng$ which contains medical contexts, queries, query-semantic information, relevant images, and extractive answers. We also propose a transformer-based architecture $M^3QAFrame$ for sentence and image relevance identification by leveraging the knowledge obtained from both modalities. Unlike previous works, our model accepts and yields both text and multiple images. We establish how indispensable context images are, as they provide unique and complementary information on top of context documents to improve MsQA. Our model relatively outperforms the existing state-of-the-art architectures for multi-span question answering as well as SOTA VLMs. In our study, we collaborated with medical specialists who found that adding images enhances the overall user experience. Doctors particularly valued images during their second-level verification process, as it gave them a better understanding of the patient’s case. In such scenarios, our model proves to be highly valuable. In the future, we intend to leverage the pre-training of open-source VLMs by fine-tuning them on our dataset and improving the modality fusion technique.

\section{Ethical Considerations}
Our project introduces a new dataset, M$^3$ QuestionIng. To achieve this, we collaborated with a medical practitioner, who is also
a co-author of this paper to ensure accuracy and
quality control throughout data collection to validation. To uphold ethical standards, we compensated all volunteers in alignment with
government's minimum wage regulations. We paid utmost attention towards privacy concerns and ensured the dataset was free of any images that could compromise individuals’ privacy. We are deeply committed to ethical principles and the responsible use of AI for social good.
\bibliographystyle{ACM-Reference-Format}
\bibliography{main}
\section{Appendix}
\textbf{Zero-Shot Prompt}

\begin{tcolorbox}[
    width=\linewidth, 
    colback=white!95!gray,
    colframe=orange!60!black,
    boxrule=0.5pt,
    arc=5pt]

\textbf{Prompt:} You are given a set of context sentences and a set of images. Your task is to identify the most relevant sentences and images that answer the question.

\textbf{CONTEXT:} \texttt{\{context\}} \\
\textbf{IMAGES:} \texttt{\{image\_set\}} \\
\textbf{QUESTION:} \texttt{\{question\}} \\
\textbf{TASK:}
\begin{itemize}
    \item List the sentences most relevant to answering the question.
    \item List the images most relevant to answering the question.
\end{itemize}

\end{tcolorbox}

\textbf{Few-Shot Prompt}
\begin{tcolorbox}[
    width=\linewidth, 
    colback=white!95!gray,
    colframe=orange!60!black,
    boxrule=0.5pt,
    arc=5pt
]
\textbf{Prompt:} You are a medical expert whose task is to analyze the given context sentences and medical images to answer the question.

\textbf{Guidance Example:} \texttt{\{example\}} \\
\textbf{Context:} \texttt{\{context\}} \\
\textbf{Images:} \texttt{\{image\_set\}} \\
\textbf{Question:} \texttt{\{question\}} \\
\textbf{Relevant Sentences:} \texttt{\{sentences\}} \\
\textbf{Relevant Images:} \texttt{\{images\}} \\

\textbf{TASK:}
For the given context: \{context\}, images: \{image\_set\}, and question: \{question\}, identify the most relevant sentences and images that answer the question as per the \{example\}.

\end{tcolorbox}

\end{document}